\RequirePackage[2020-02-02]{latexrelease}
\documentclass[superscriptaddress,preprint,longbibliography,aps,nofootinbib,showpacs,12pt]{revtex4}

\usepackage{graphicx}  
\usepackage{bm}        
\usepackage{amssymb}   
\usepackage{epsfig}
\usepackage{amsmath,amsfonts,times}
\usepackage{bm}

\usepackage{epstopdf}
\usepackage[caption=false]{subfig}
\usepackage{natbib}
\usepackage{soul}
\usepackage[utf8x]{inputenc}

\begin{document}

\title{Geometrical aspects of the multicritical phase diagrams for the Blume-Emery-Griffiths model} %

\author{Nigar Alata} 
\affiliation{Institute of Science, Akdeniz University, 07058 Antalya, Türkiye}
\affiliation{Food Safety and Agricultural Research Centre, Akdeniz University, 07058 Antalya, Türkiye}

\author{R{\i}za Erdem}
\email{e-mails: rerdem@akdeniz.edu.tr; rizaerdem07@gmail.com}
\affiliation{Department of Physics, Akdeniz University, 07058 Antalya, Türkiye}

\author{G\"{u}l G\"{u}lp{\i}nar }
\affiliation{Department of Physics, Dokuz Eyl\"{u}l University, 35210 \.{I}zmir, Türkiye}

\begin{abstract}

As a continuation of our preceding work [R. Erdem and N. Alata, \emph{Eur. Phys. J. Plus} {\bf 135}, 911 (2020), https://doi.org/10.1140/epjp/s13360-020-00934-3], we used the thermodynamic geometry in the Ruppeiner formalism to study the geometrical aspects of the multicritical phase diagrams for the spin-1 Blume-Emery-Griffiths model in the presence of crystal field. We derived an expression for the thermodynamic curvature or Ricci scalar ($R$) and analyzed its temperature and crystal field behaviours near interesting critical and multicritical points. Our findings are presented as geometrical phase diagrams including critical and multicritical topology. From these diagrams, new vanishing curvature lines ($R=0$) extending into the ferromagnetic or paramagnetic phases beyond the critical points and zero point temperature are observed.
\end{abstract}

\pacs{05.70.a, 05.70.Fh, 75.10.Hk, 02.40.Ky}

\date{\today}
	
\maketitle

\section{Introduction}

The Blume-Emery-Griffiths (BEG) model is a spin-1 Ising system with bilinear exchange energy constant $J$, biquadratic exchange energy constant $K$ and external magnetic field $H$ in which a single-ion anisotropy parameter $D$ is included. It was originally introduced \cite{[1]} in order to explain the phase separation and superfluidity in the $^3He$ - $^4He$ mixtures, as well as to describe other multi-component physical systems, such as liquid-crystal mixtures, semiconductors, microemulsions, ... etc. The properties of this model were established using various techniques \cite{[2],[3],[4],[5],[6],[7],[8],[9],[10],[11],[12],[13],[14],[15],[16]}. As a novel analysis, we have studied in our recent paper \cite{[17]} referred as paper I in the following, the thermodynamic geometry of the spin-1 BEG model in the absence of single-ion anisotropy ($D=0$), also known as the isotropic BEG model. Based on a metric in a two-dimensional ($2\mathcal{D}$) order parameter state space, we derived an expression for the thermodynamic curvature or Ricci scalar ($R$) in paper I and investigated temperature and magnetic field dependence of $R$ for various values of bilinear to biquadratic ratio. Besides the divergence singularity and finite jumps connected with the phase transitions we have found field-dependent broad extrema in the Ricci scalar.

In this work, we extend our previous geometrical analysis to the spin-1 BEG model including single-ion anisotropy ($D\neq0$) named as anisotropic BEG model or also BEG model in short. Compared with the isotropic case, it has very rich phase diagrams which have critical behaviours including multicritical points \cite{[1], [2],[3],[4],[5],[6],[7],[8],[9],[10],[11],[12],[13],[14],[15],[16]}. Although such phase diagrams  have been of considerable interest for years, no attention has been devoted to the study of geometrical phase transitions undergoing in the same phase diagrams. Using Ruppeiner’s conjecture Sanwari and Sahay have recently analyzed in detail the geometry of phase coexistence in \cite{[18], [19]}. Different from their works, we have investigated the critical and multicritical behaviours of $R$ by choosing $2\mathcal{D}$ order parameter metric and studied the geometrical phase diagrams (GPDs) including vanishing curvature lines as well as the $R$ extrema curves in the multicritical topology.

Outline of this work is as follows. In Sec. II, we shortly describe the Hamiltonian of the spin model and show Gibbs free energy with its minimization yielding the self-consistent equations. In Sec. III, we give an overview for the thermodynamic geometry under Ruppeiner formalism. In order to show the details of the theoretical framework we present the derivations of the Christoffel symbols, curvature tensor components, Ricci tensor components and Ricci scalar for the BEG model. In Sec. IV, we investigate the behavior of $R$ from the viewpoint of critical and multicritical phenomena. Furthermore, we obtain the GPDs for four phase diagram topologies. We finally conclude all our findings in Sec. V.

\section{Model hamiltonian, Gibbs free energy and Multicritical phase diagrams}

The Hamiltonian of the BEG model is given by \cite{[1],[2],[3],[4],[5],[6],[7],[8],[9],[10],[11],[12],[13],[14],[15],[16]}
\begin{equation}
	\mathcal{H}= - J \sum_{<ij>} S_i S_j - K \sum_{<ij>} S_i^2 S_j^2 + D \sum_{i}^N S_i^2 - H \sum_{i}^N S_i, \label{ham}
\end{equation}  
where spin variable $S_i=\pm1, 0$ at each lattice $i$, $<ij>$ indicates summation over all pairs of nearest-neighbour sites and $N$ is the number of lattice sites. The properties of the system at equilibrium are generally determined self-consistently using Helmholtz free energy calculations. But, as it will be seen in the next section, in order to determine a metric we use Gibbs free energy which is better suited thermodynamic potential for the 2$\mathcal{D}$ order parameter state space \cite{[17]}. As a starting point, we now briefly write a simple expression for the Gibbs free energy as
\begin{equation}
	G = U - TS - HM + DQ,   
\end{equation}
where $U$, $T$, $S$, $M$ and $Q$ are called the internal energy, absolute temperature, entropy, dipolar order and quadrupolar order, respectively. Here $S=N\sigma$ ($\sigma$ entropy per site), $M=Nm$ and $Q=Nq$. The order parameters $m$ and $q$ are the average magnetization given by $m \equiv <S_i>$ and the average of squared magnetization given by $q \equiv <S_i^2> $, where $<...>$ denotes the thermal expectation value. Using mean-field approximation in the Bragg-Williams formalism, the Gibbs free energy can be rewritten in the form    
\begin{equation}
	\phi( m, q) =  \frac{1}{\gamma J}\frac{G}{N} = - \frac{1}{2} m^2 - \frac{1}{2}rq^2 + \theta \sum_{i=1}^{3} p_{i}\ln p_{i}- h m + dq, \label{f-mq}
\end{equation}
where $r=K/\gamma J$ ($\gamma$ is the lattice coordination number or number of nearest neighbour sites) is the coupling ratio constant, $\theta=k_B T/\gamma J$  is the reduced temperature ($k_B$ the Boltzmann constant), $h = H/\gamma J$ is the reduced magnetic field, $d=D/\gamma J$ is the reduced crystal field, and $p_{i}$ are the probabilities of the spin states which can be expressed as linear combinations of the order parameters
\begin{equation}
	p_{1}=\frac{1}{2}(m+q),\,\,\, p_{2}=1-q,  \,\,\, 	p_{3}=-\frac{1}{2}(m-q).  \label{omega-mq}
\end{equation}
For the system at equilibrium, dipolar and quadrupolar order parameters are obtained self-consistently by a minimization of Eq. (3) with respect to $m$ and $q$ (i.e., $\partial \phi / \partial m = 0$, $\partial \phi / \partial q = 0$):
\begin{equation}
	m = \frac{ 2  \sinh(m/\theta)}{\exp[(d-rq)/\theta] +2 \cosh (m/\theta)}, \,\,  	q = \frac{ 2  \cosh(m/\theta)}{\exp[(d-rq)/\theta] +2 \cosh (m/\theta)}.   \label{m-q}
\end{equation}
Evidently, note that $h=0$ and then the thermodynamic behaviours of the system and phase transitions between the ferromagnetic (F) and paramagnetic (P,P1,P2) phases depend on $r,d$ and $\theta$ in Eq. (5). From the numerical solutions of these equations, equilibrium states and the relevant phase transitions with multicritical phenomena are previously investigated in \cite{[13],[16]}. For the convenience of our later discussions, using Maxwell construction law (equal area principle), we reobtained four diagrams which were previously presented by Hoston and Berker in \cite{[4],[5]} and replotted the results into geometrical phase diagrams in the next section. We can summarize the significant properties of these diagrams as follows: for the attractive biquadratic interaction with $r=3$, a first- and second-order phase transitions take place between F and P2 phases while only the first-order transition occurs across the F-P1 and P1-P2 phase coexistence lines. In this phase diagram topology, a tricritical point (TCP), a triple point (TP) and an external critical point (C) exist. On the other hand, the system exhibits reentrant behaviours for the repulsive biquadratic interactions. For example, the phase diagram develops a doubly reentrant (F-P-F-P) topology for $r=-0.15$ where a second-order transition line terminates at the TCP. For $r=-0.5$, besides single reentrance phenomenology (P-F-P) there occurs a critical end point (CEP) and, inside the ferromagnetic phase, a first order line part terminating at an internal critical point C$'$. As $r$ approaches -1, above structure collapses toward zero temperature and thus CEP no longer exists for $r=-1$. At the end, for $r=-1$, a singly reentrant critical line reaches zero temperature point (Z), which, as a critical point characterized by fluctuations at zero temperature. Detailed mean field investigations of BEG model by Hoston and Berker revealed the fact that there exists two additional ordered phases (ferrimagnetic and antiquadrupolar) and three distinct phase diagram topologies for $r<-1$. As mentioned above since we have chosen 2$\mathcal{D}$ order parameter manifold in this study we have focused our calculations on $r\geqslant-1$ case.

\section{Termodynamic geometry of the BEG model in the Ruppeiner formalism}

In this section, we start with a brief review of the steps in the thermodynamic geometry. Firstly, an $n$-dimensional thermodynamic state space or Riemannian manifold ($\mathcal{M}$) is defined. A metric which is the square of distance between very closely two points in this space is given by
\begin{equation}
	ds^2 = g_{ij} dx^i dx^j, \qquad ( i,j = 1,2,...,n), \label{metric}
\end{equation}
where $x^i$ denotes the various thermodynamic parameters. In this case, components of metric tensor which are evaluated in the state of minimum free energy can be expressed using the Ruppeiner approach as follows \cite{[20],[21]} 
\begin{equation}
	g_{i j} = -\beta \partial_i \partial_j \phi\ ,\label{metriccomponet}
\end{equation}
where $\beta = 1/k_B T$, $\partial_i = \partial / \partial x^i$ and $\phi$ is the thermodynamic potential per site. Using the above definition, one can obtain the Christoffel symbols as 
\begin{equation}
	\Gamma^{i}_{jk} = \frac{1}{2}g^{il}(\partial_k g_{lj}+\partial_j g_{lk}-\partial_l g_{jk}).\label{metriccomponet}
\end{equation}
A calculation by the above Christoffel symbols gives the Riemann and Ricci tensors: 
\begin{equation}
	R^{i}_{jkl} = \partial_k \Gamma^{i}_{jl}-\partial_l \Gamma^{i}_{jk}+ \Gamma^{i}_{mk} \Gamma^{m}_{jl}-\Gamma^{i}_{ml} \Gamma^{m}_{jk},\label{metriccomponet}
\end{equation}
\begin{eqnarray}
	& & R_{ij} = R_{inj}^{n},\qquad  \label{einstein}
\end{eqnarray}
where we are using the Einstein summation in the same upper and lower indexes. Then, one can write the Ricci scalar or curvature scalar expression, such that
\begin{eqnarray}
	& & R = g^{ij}R_{ij}, \qquad  \label{ricci}
\end{eqnarray}
with $g^{ij}$ in Eq. (11) being the components of contravariant metric tensor. The quantity $R$ provides important information about the microscale properties of thermodynamic systems. Not only it measures the complexity of a system, but also indicates information about effectively attractive ($R<0$) or repulsive ($R>0$) nature of the particles or spins of the system around phase transition points. In addition, $R=0$ (or vanishing curvature) line seperates the thermodynamic phase regions into $R<0$ and $R>0$ phases. Based on this interpretation of Eq. (11), a number of magnetic and fluid systems has been analyzed theoretically for the last decade [22-34].

As in the previous paper \cite{[17]}, for the spin model under consideration, we also define a 2$\mathcal{D}$ order parameter manifold by $( x^1,x^2) = (m,q)$. Then, the components of a nondiagonal covariant metric tensor can be obtained with the help of Eqs. (3) and (7) 
\begin{equation}
	g_{11} = \frac{m^2 + q\theta - q^2 }{\theta(m^2 - q^2)}, \,\, g_{12} = g_{21} = -\frac{m}{m^2 - q^2}, \,\, g_{22} = - \frac{rq^3-rq^2+(-rm^2+\theta)q-m^2(\theta-r)}{\theta(q - 1)(m^2 - q^2)}, \label{metric-coef}
\end{equation}
where the derivatives are evaluated in the equilibrium state. From the above components and definition (8), the Christoffel symbols are evaluated as follows:
\begin{equation}
	\Gamma^{1}_{11} = - \frac {m\theta[rq^2 +( \frac{1}{2}\theta-r)q+\frac{1}{2}\theta]} {\mathcal{A}(q+m)(q-m)} \, ,\label{metriccomponet}
\end{equation}

\begin{equation}
	\Gamma^{1}_{12} = \Gamma^{1}_{21} = \frac{1}{2} \frac {\theta[q^3r-rq^2+(m^2r+\theta)q+m^2(\theta-r)]} {\mathcal{A}(q+m)(q-m)}\, ,\label{metriccomponet}
\end{equation}

\begin{equation}
	\Gamma^{1}_{22} =  - \frac{1}{2} \frac {m\theta(2q^3r-4q^2r+m^2\theta+2qr-\theta)} {\mathcal{A}(q - 1)(q+m)(q-m)} \, ,\label{metriccomponet}
\end{equation}

\begin{equation}
	\Gamma^{2}_{11} =  \frac{1}{2} \frac {\theta(q - 1)(q^2-q\theta+m^2)} {\mathcal{A}(q+m)(q-m)} \, ,\label{metriccomponet}
\end{equation}

\begin{equation}
	\Gamma^{2}_{12} = \Gamma^{2}_{21} = - \frac {m\theta(q - \frac{1}{2}\theta)(q-1)} {\mathcal{A}(q+m)(q-m)} \, ,\label{metriccomponet}
\end{equation} 

\begin{equation}
	\Gamma^{2}_{22} = - \frac {\mathcal{B}\theta} {\mathcal{A}(q - 1)(q+m)(q-m)}, \label{metriccomponet}
\end{equation} 

where the coefficients $\mathcal{A}$ and $\mathcal{B}$ are defined by
\begin{eqnarray}
	& &	\mathcal{A} = q^3r-r(\theta+1)q^2+[(r+1)\theta-m^2r]q-m^2\theta+m^2r-\theta^2,\label{metriccomponet}
\end{eqnarray}  

\begin{eqnarray}
	& &	\mathcal{B} = q^3-\bigg(\frac{3}{2}m^2+\theta+\frac{1}{2}\bigg)q^2+\bigg[\bigg(\frac{1}{2}\theta+1\bigg)m^2+\frac{1}{2}\theta\bigg]q+\frac{1}{2}m^4-\frac{1}{2}m^2.\label{metriccomponet}
\end{eqnarray} 

Using (13)-(18) in (9) non-zero curvature tensor components are found as

\begin{equation}
	R^{2}_{121} =  \frac{1}{4} \frac{\theta^2 \mathcal{C}(q^2-q\theta-m^2)}{\mathcal{A}^2(q+m)(q-m)} \, ,\label{metriccomponet}
\end{equation}

\begin{equation}
	R^{1}_{112} = R^{2}_{221} =  \frac{1}{4} \frac{m\theta^3 \mathcal{C}}{\mathcal{A}^2(q+m)(q-m)} \, ,\label{metriccomponet}
\end{equation}

\begin{equation}
	R^{1}_{212} =  \frac{1}{4} \frac{\theta^2\mathcal{C}[q^3r-rq^2+(\theta-m^2r)q-m^2(\theta-r)]}{\mathcal{A}^2(q - 1)(q+m)(q-m)} \, ,\label{metriccomponet}
\end{equation}

and using (21)-(23) in (10) Ricci tensor components

\begin{equation}
	R_{11} =   \frac{1}{4} \frac{\theta^2 \mathcal{C}(q^2-q\theta-m^2)}{\mathcal{A}^2(q+m)(q-m)} \, ,\label{metriccomponet}
\end{equation}

\begin{equation}
	R_{12} = R_{21} =  \frac{1}{4} \frac{m\theta^3 \mathcal{C}}{\mathcal{A}^2(q+m)(q-m)} \, ,\label{metriccomponet}
\end{equation}

\begin{equation}
	R_{22} =   \frac{1}{4} \frac{\theta^2\mathcal{C}\big[q^3r-rq^2+(\theta-m^2r)q-m^2(\theta-r)\big]}{\mathcal{A}^2(q - 1)(q+m)(q-m)},\label{metriccomponet}
\end{equation}
where $\mathcal{C}$ is given by

\begin{equation}
	\mathcal{C} = rq^2+(2-2r)q+m^2-\theta+r-1. \label{metriccomponet}
\end{equation}
Finally, based on Eq. (\ref{ricci}) Ricci scalar is obtained in terms of the equilibrium values of $m$ and $q$ as follows:

\begin{equation}
	R (m,q) = - \frac{1}{2} \frac{\mathcal{C} \theta^3} {\mathcal{A}^2}. \label{ricci-2}
\end{equation}
In the next section, we will give our graphical results of numerical calculations on $R$ and discuss the findings. Understandably, we will clarify in the zero magnetic field the phase transition behaviours of $R$, particularly curvature singularity at the points TCP, TP, C,  C$'$, CEP, Z.

\section{Results and discussion}

We now show and discuss some features of Eq. (28). In this context, Fig. 1 displays both reduced temperature and reduced single-ion anisotropy dependence of the curvature scalar $R$ for the case of attractive ($r=3$) biquadratic interactions. Fig. 1a-c represent temperature variation of $R$ for $d=d_{TP}=2.0$, $d=d_C=2.02$ and $d>d_C$ cases, respectively. Vertical dotted lines indicate the points of TP and C in Figs. 1a, b, respectively, while in Fig. 1c they refer to the temperature where $R$ changes sign. $R=0$ case is denoted by a horizontal dotted line. As can be seen from Fig. 1a, along the F-P1 phase boundary line with $d_{TP}=2.0$, $R$ takes negative values. While $\theta_{TP}$ is approached  $R$ decreases rapidly, but it suddenly starts to take positive values when the $\theta_{TP}$ is reached, namely $R>0$ in P2 phase. Likewise, in Figs. 1b and 1c, we represent the results for $d_C\geqslant2.02$ to understand the sign of $R$ in the P1 phase. Different from the case at the TP, a smooth transition starts to occur from P1 phase with $R<0$ to P2 phase with $R>0$ at C (Fig. 1b). Here, we also show a sharp minimum ($R_M$) in the P1 phase side just below $\theta_{C}$ and a sharp maximum ($R_P$) in the P2 phase side slightly above $\theta_{C}$. As is appearent from Fig. 1c, these extrema of $R$ decrease and shift to higher temperatures while becoming broad with the increases of $d$. Similarly, in Fig. 1d-f, calculations are given in the ($d,R$) plane to show single ion anisotropy dependence of $R$ for temperatures $\theta<\theta_{TP}$, $\theta_{TP}<\theta<\theta_{C}$ and $\theta_{C}<\theta_{TCP}<\theta$, respectively. For all values of $\theta$, the quantity $R$ makes sharp cusps in the $R<0$ region which become more sharply as the temperature is raised at TP point (Fig. 1d). In Fig. 1e, for a chosen value of $\theta$ which is between $\theta_{TP}$ and $\theta_C$, $R$ changes its sign suddenly while occuring F-P2-P1 first-order phase transitions, i.e., $R<0\rightarrow R>0\rightarrow R<0$. In addition, we observed a $R_M$ in the P2 phase with $R>0$. From Fig. 1f, we observed in the case of $\theta>\theta_{TCP}$, the following changes in $R$ while $d$ increases: (1) $R$ decreases in F phase ($R<0$) and makes a minimum, (2) it starts to increase, changes its sign and diverges to positive infinity ($R\rightarrow\infty$) around the F-P2 phase transition line, (3) it decreases in P2 phase ($R>0$) and a new minimum is formed, (4) starts to increase again and reaches a maximum, (5) it begins get down to, changes its sign ($R<0$) and goes on decreasing in P1 phase, (6) finally, reaches a minimum in P1 phase and goes to $R=0$ line with an increase again. \begin{figure*}
	\centering
	\includegraphics[width=0.42\linewidth]{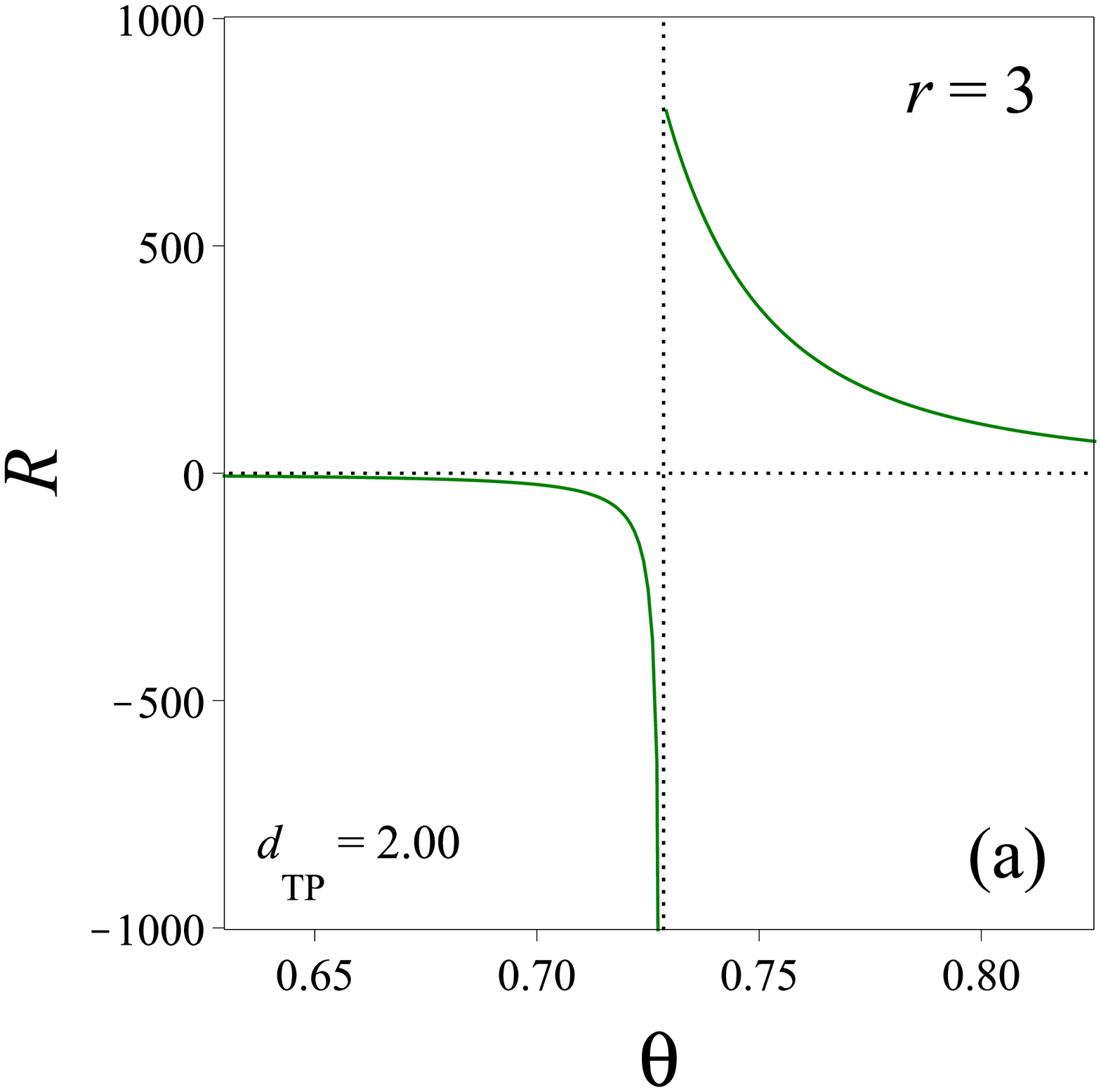}
	\includegraphics[width=0.42\linewidth]{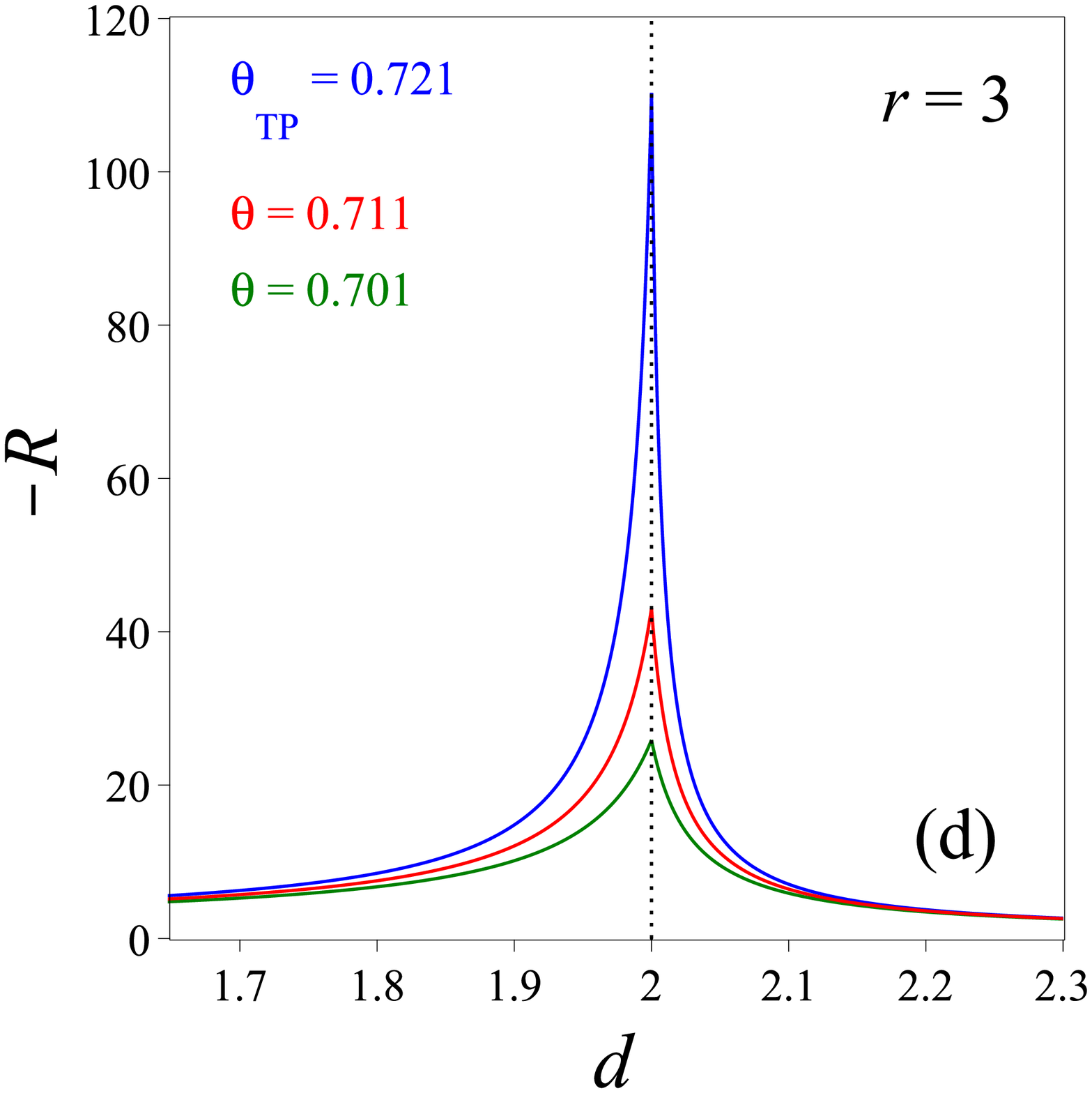}
	\includegraphics[width=0.42\linewidth]{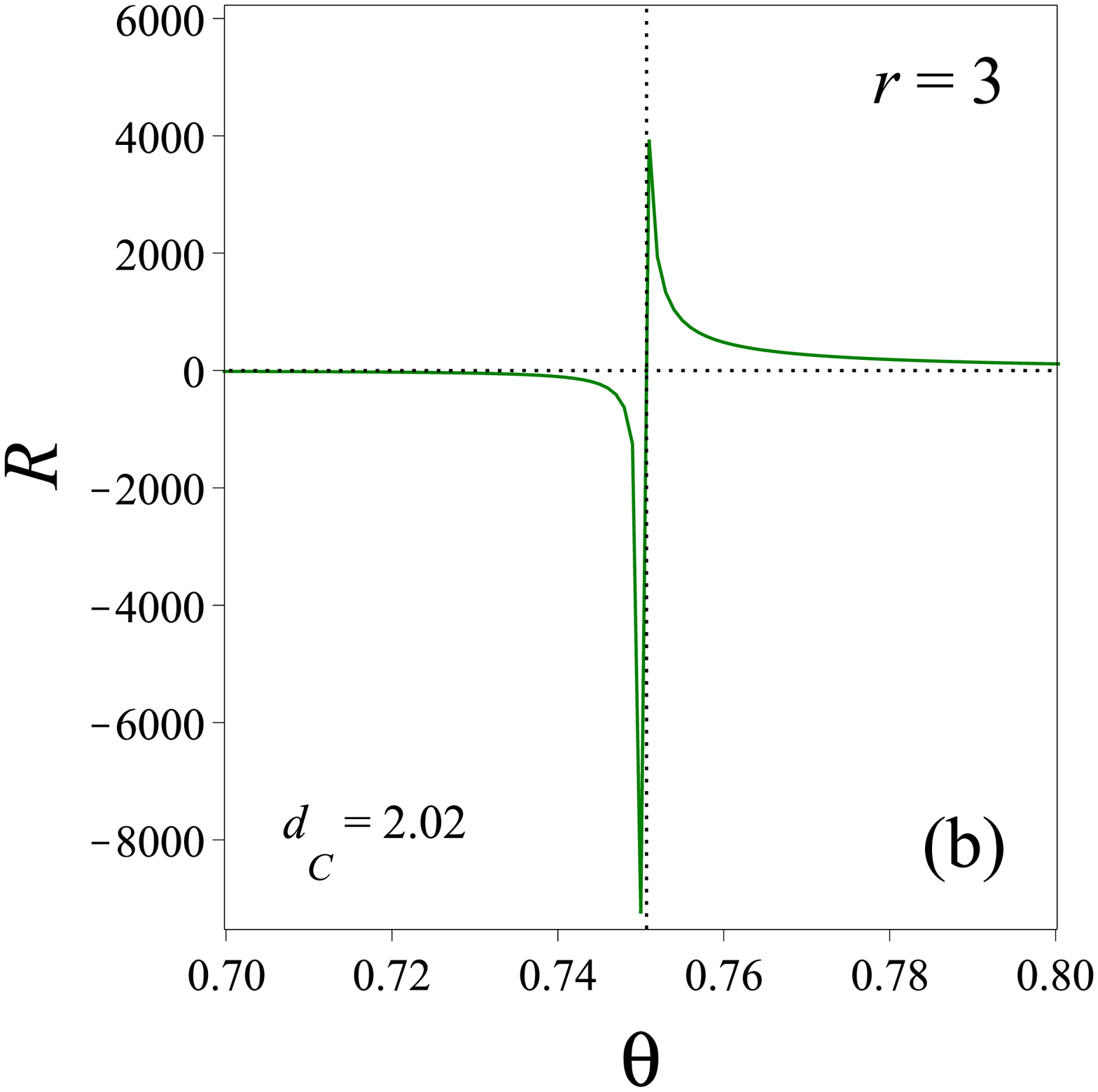}	
	\includegraphics[width=0.42\linewidth]{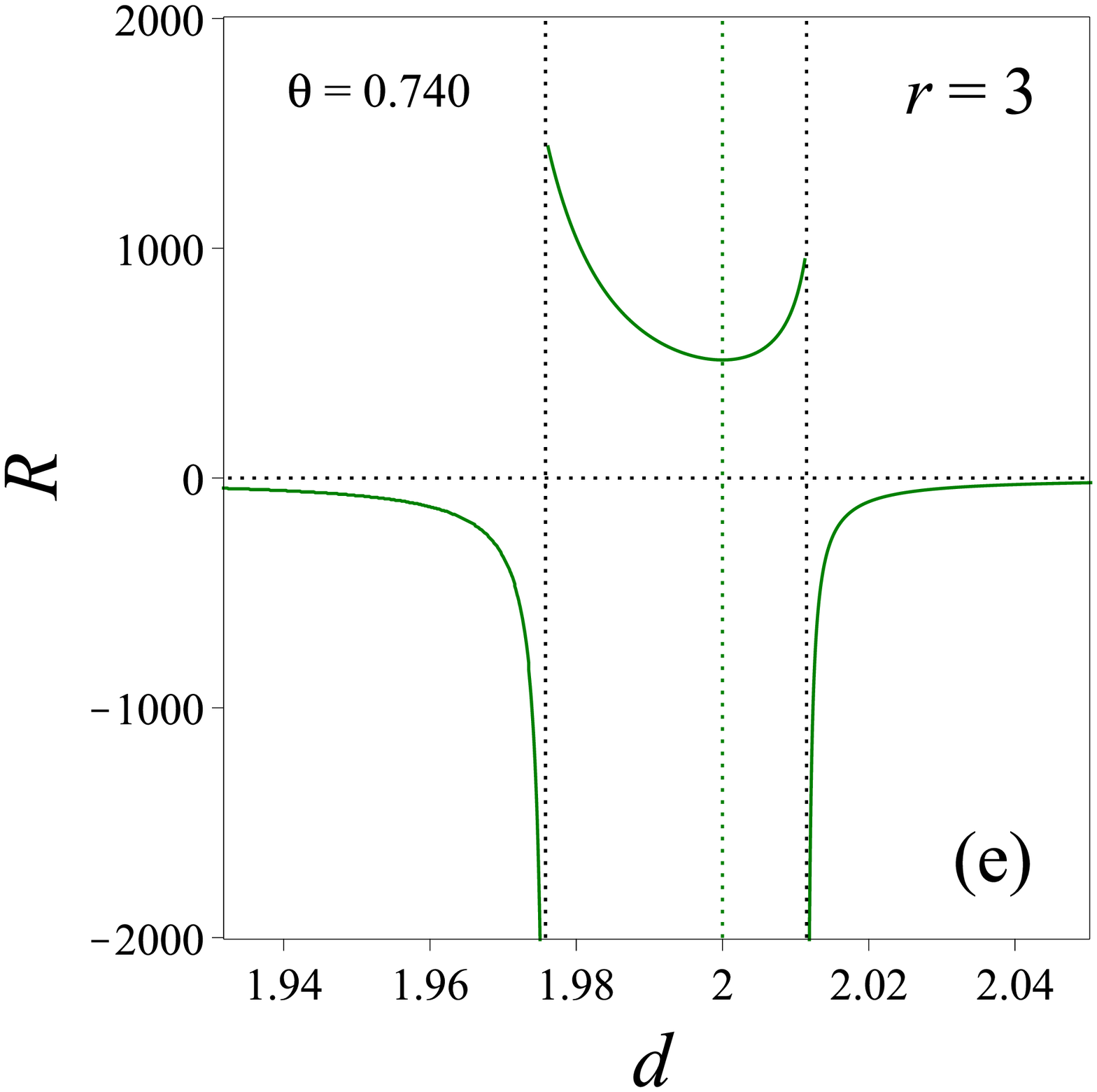}
	\includegraphics[width=0.42\linewidth]{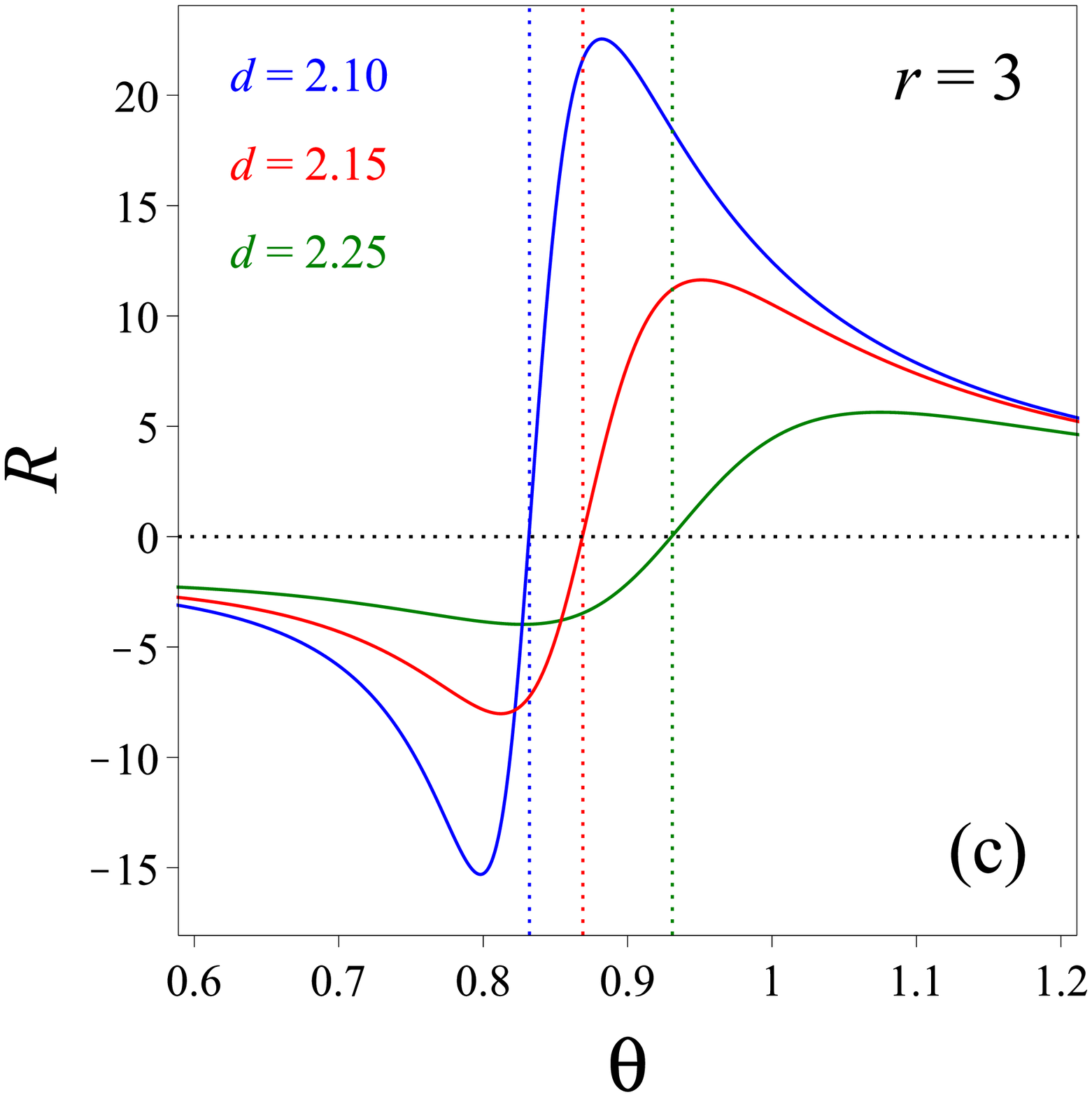}
	\includegraphics[width=0.42\linewidth]{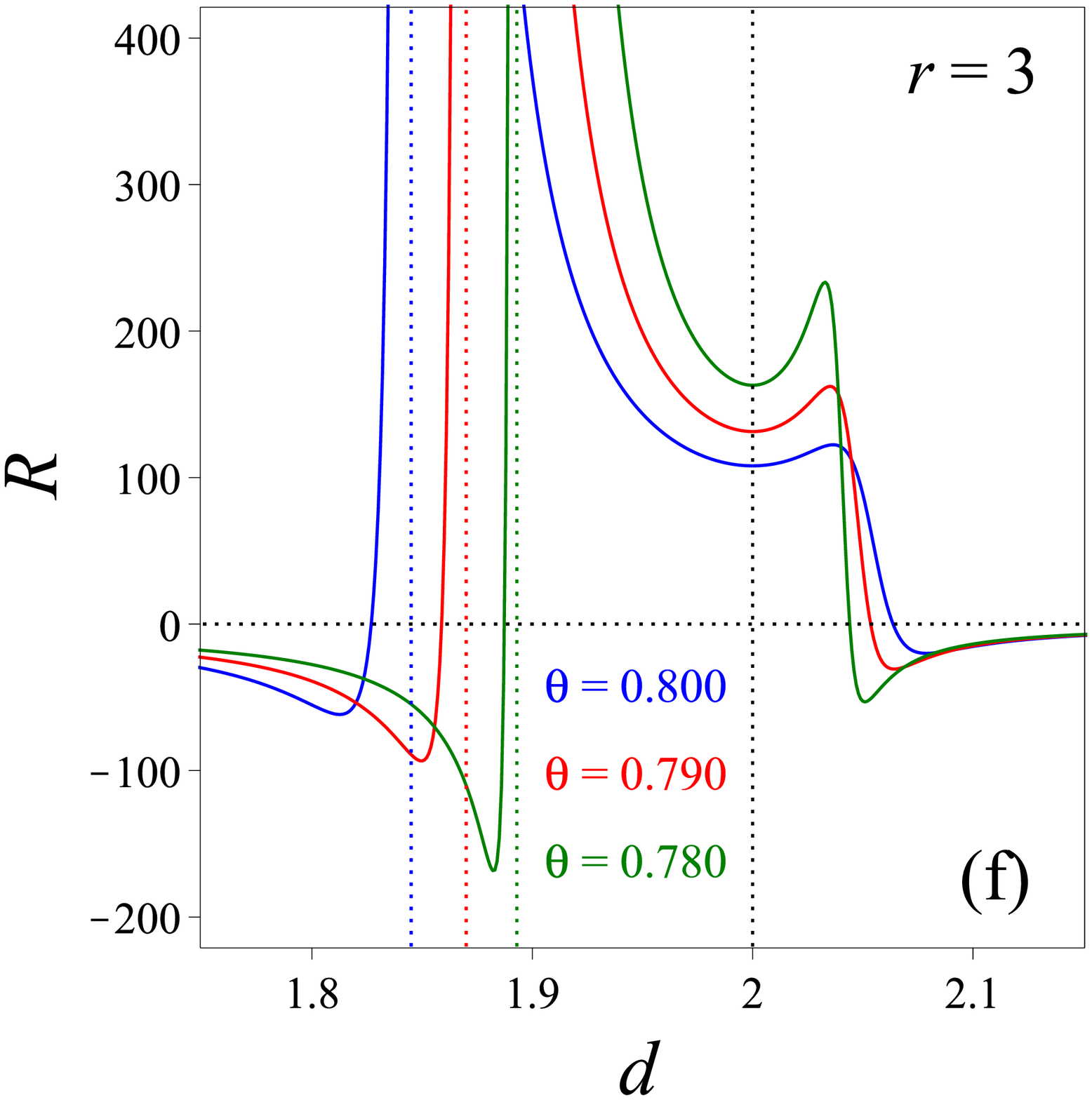}
	\caption{{\textbf{a-c}} Ricci scalar $R$ vs. reduced temperature $\theta$ for different values of $d$ in the $\theta-R$ plane, {\textbf{d-f}} Ricci scalar $R$ vs. reduced single ion anisotropy $d$ for different values of $\theta$ in the $d-R$ plane, for $r=3$. Vertical dotted lines correspond to the phase transition points, the temperature values at which $R$ changes sign and the minimum in P2 phase.}
	\label{fig1}
\end{figure*}

We performed, in Fig. 2, similar calculations for the repulsive biquadratic interaction with $r=-0.5$. Now, curvature scalar $R$ starts to increase independent of $d$ values from negative infinity ($-\infty$) in the F phase ( $d<0.25$). After making a maximum $R_P$ and a $d$-dependent minimum $R_M$, respectively, it begins to rise again in the same phase (Fig. 2a). $R$ changes its sign while increases continuously in this phase and reaches a F-P second-order phase transition point where it diverges to positive infinitiy ($R\rightarrow\infty$). In the P phase, $R$ starts to decrease from $+\infty$ while around the critical temperature and remains always positive ($R>0$) in this phase. Different from the case in Fig. 1, the F-P-F reentrant behaviours of $R$ are observed for $d=0.3$ in Fig. 2b. In this case, $R$ makes firstly a minimum while decreases from a saturated value. It starts to raise and diverges to positive infinity around the P-F phase transition. There is a continuous drop which terminates at a minimum value in the F phase. Another tendency to $+\infty$ occurs at the F-P transition point. Finally, it decreases in the P phase region and stays always positive. Similarly, we can see the single ion anisotropy dependence of $R$ in Figs. 2c, d. As is apparent clearly from Fig. 2c, it is noticed that $R$ decreases in the F phase (or $R<0$ region) and changes its sign to $R>0$ at C$'$ point suddenly. In the P phase region, it decreases again and makes a minimum, of course, $R$ diverges to positive infinity on both sides of the F-P phase transition. Moreover, it goes on decreasing towards $R=0$ line in P phase region without changing its sign. Finally, as can be seen in Fig. 1d, there is a large jump to $R>0$ from $R<0$ at the critical end point. \begin{figure*}
	\centering
	\includegraphics[width=0.45\linewidth]{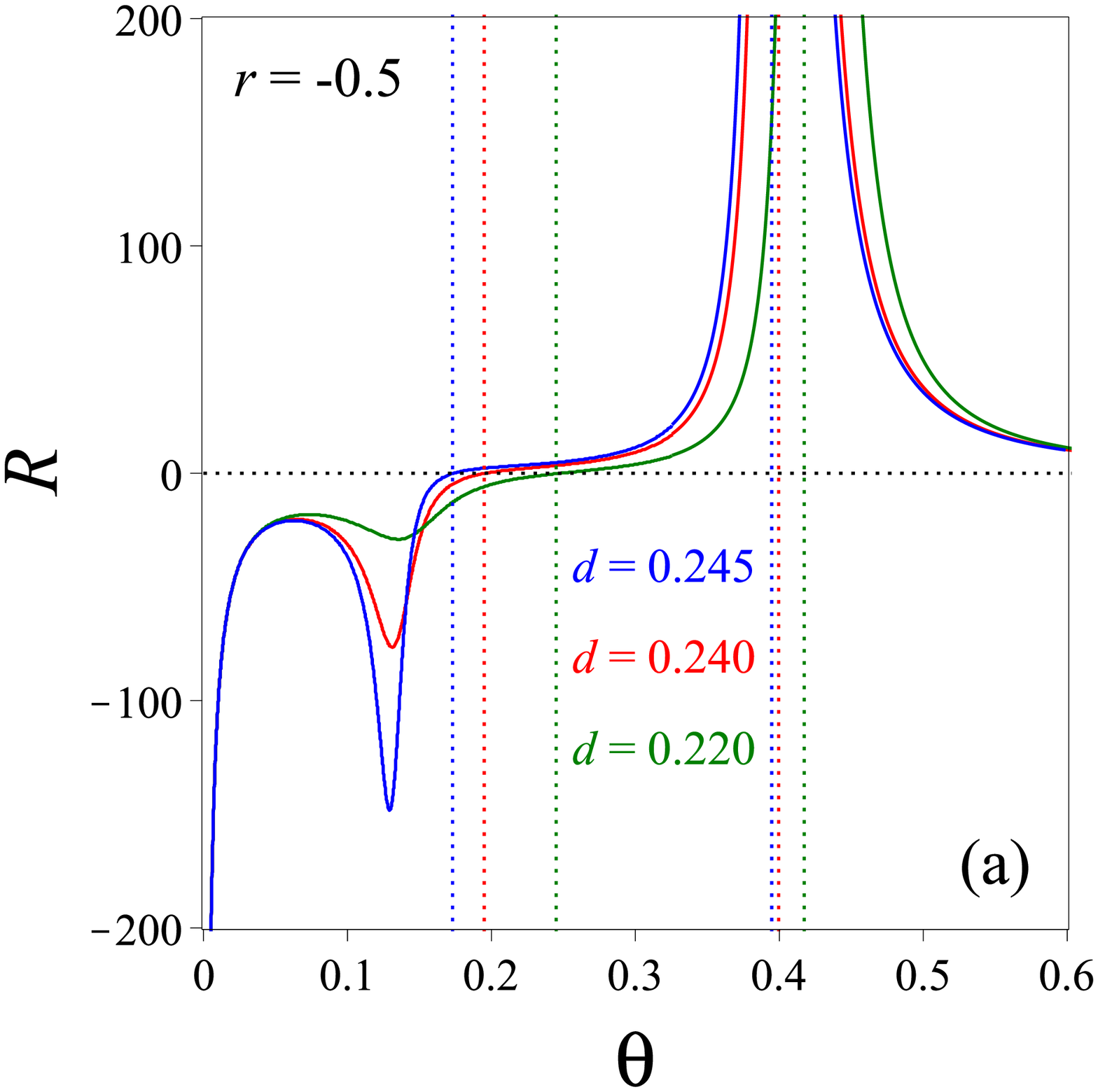}
	\includegraphics[width=0.45\linewidth]{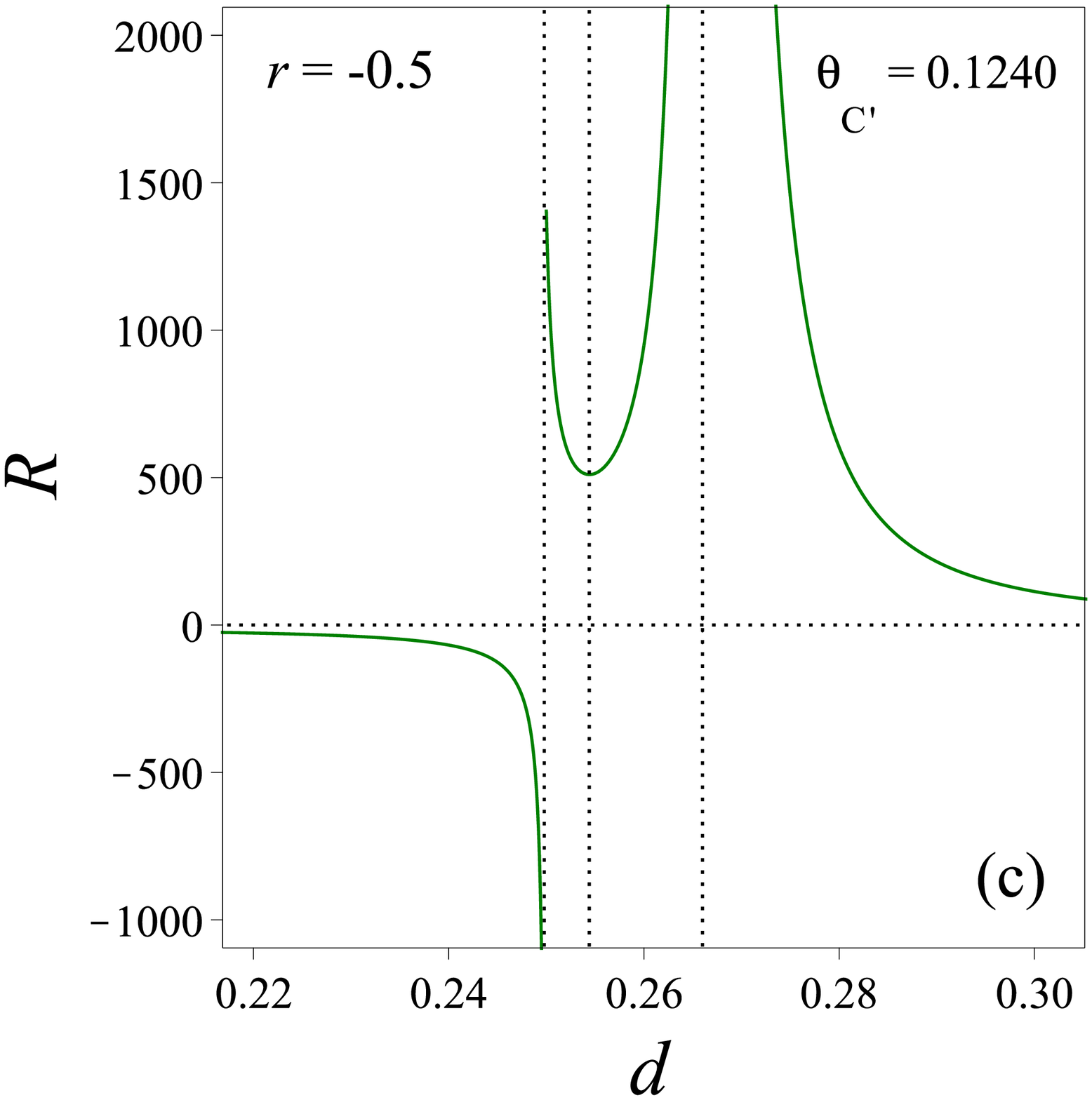}
	\includegraphics[width=0.45\linewidth]{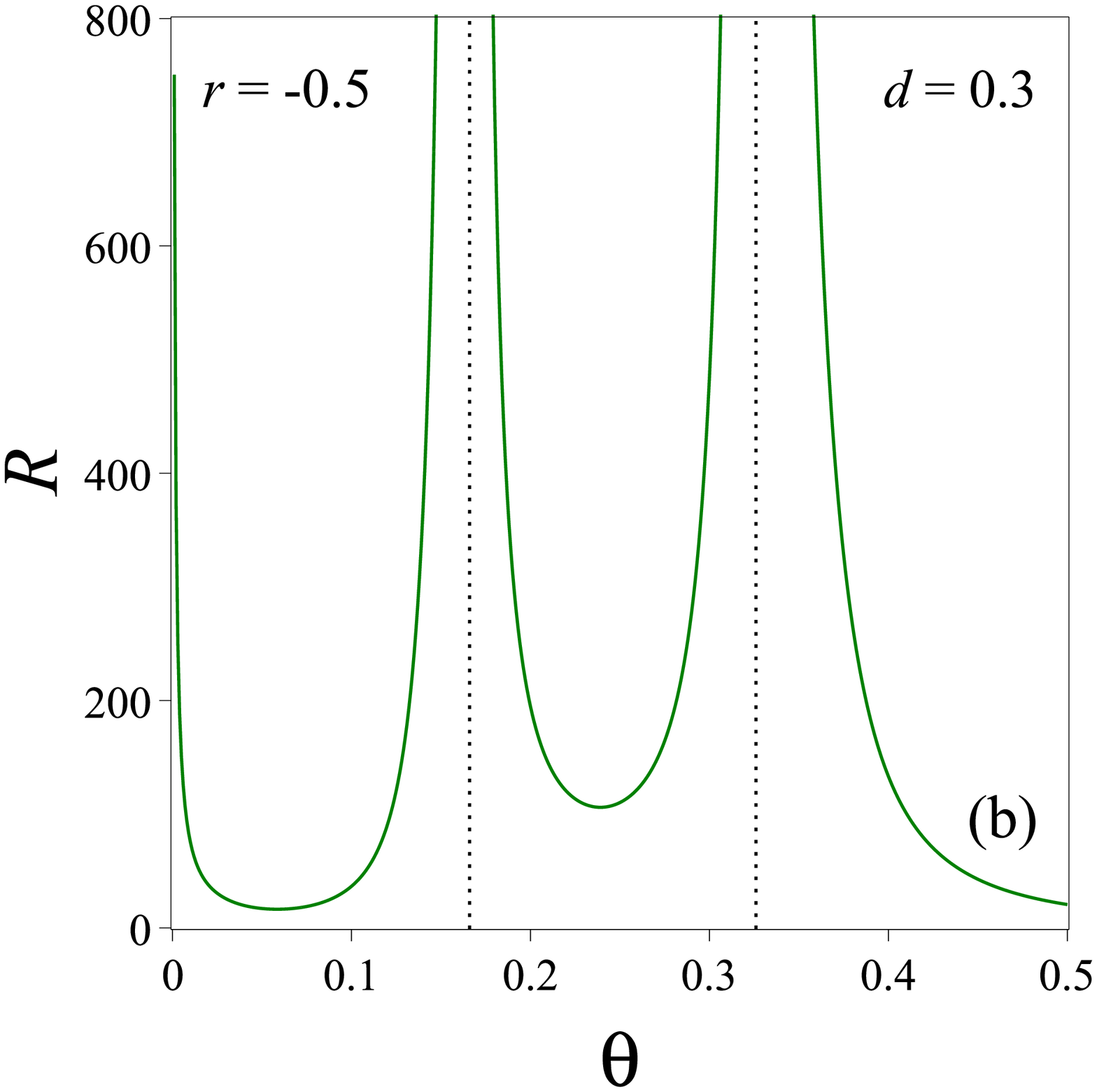}
	\includegraphics[width=0.45\linewidth]{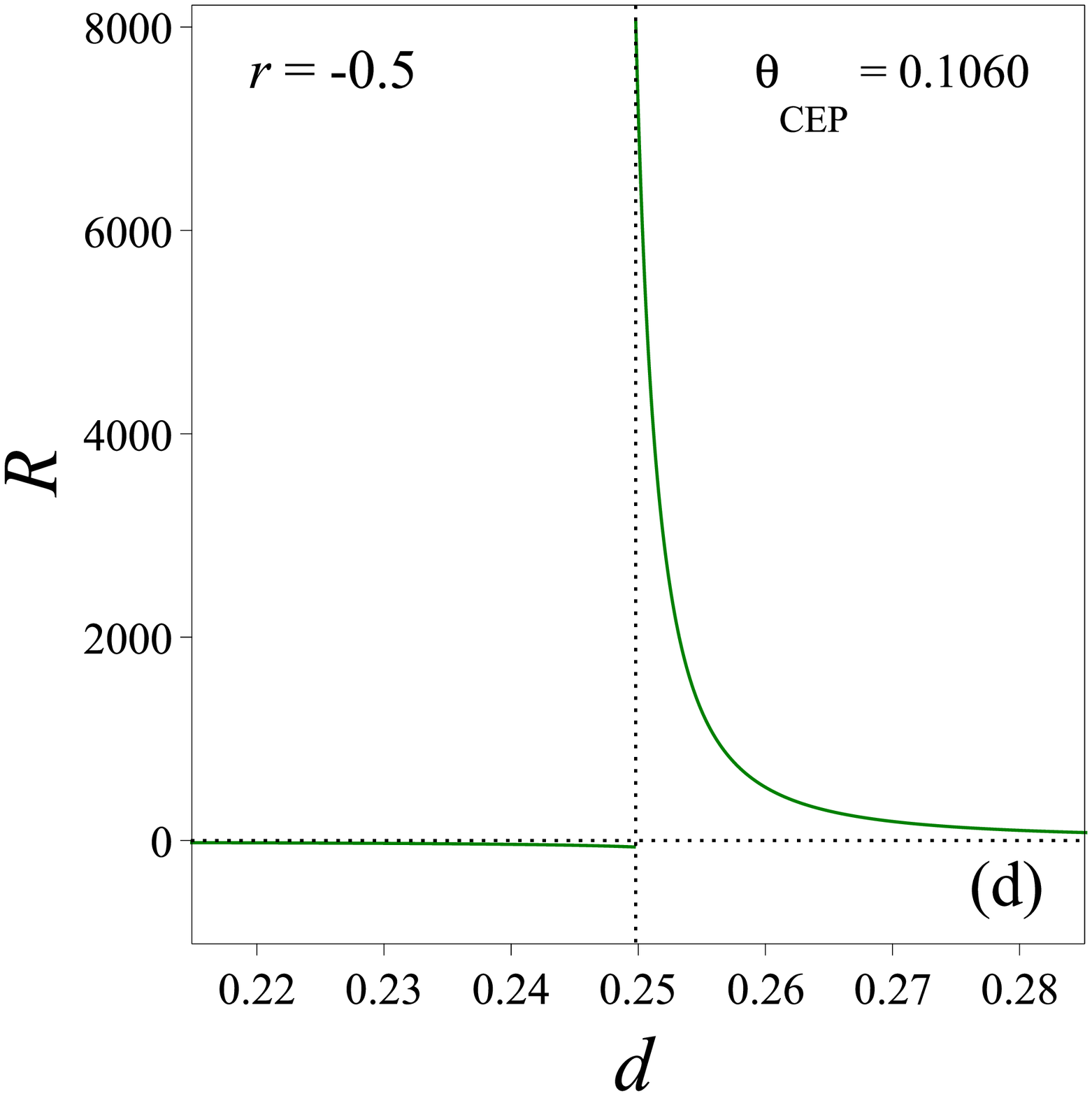}
	\caption{{\textbf{a-b}} Ricci scalar $R$ vs. reduced temperature $\theta$ for different values of $d$ in the $\theta-R$ plane, {\textbf{c-d}} Ricci scalar $R$ vs. reduced single ion anisotropy $d$ for two different values of $\theta$ in the $d-R$ plane, for $r=-0.5$. Vertical dotted lines correspond to the phase transition points and the temperature values at which $R$ changes sign.} \label{fig2}
\end{figure*}

After calculating $R$, we have obtained the GPDs which include $R>0$ and $R<0$ zones as in [17] for various $r$ values with attractive and repulsive interactions by the addition of $r=-0.15$ and $r=-1$ calculations as well. The locations of the $R=0$, $R_P$ and $R_M$ curves, which are shown by red solid lines, green solid lines and green dashed lines, respectively, were determined in the same diagrams in the ($d,\theta$) plane. It is evident from Fig. 3a that $R=0$ and $R_M$ curves observed inside F phase terminate together at a point (E) ($\theta_E=0.762, d_E=1.930$) which is on the first-order transition line, beyond the TCP. This result is similar to the geometric phase diagram of the isotropic model in paper I. Another $R_M$ curve appears as a prolongation of the P1-P2 phase coexistence line above the TP ($d_{TP}=2$) in P2 phase. Firstly, this curve extends into the region $R>0$ as a line as $\theta$ rises, then it turns to left after a specific temperature value is reached. Moreover, in this phase, an $R_P$ curve emerges which is like an extension of C. Furthermore, other $R=0$ and $R_M$ curves start again from the same critical point C ($d_C=2.02$) and extend into the P1 phase zone. In summary, according to Fig. 3a, always $R>0$ in P2 phase while partially $R<0$ in F and P1 phases.\begin{figure*}
	\centering
	\includegraphics[width=0.45\linewidth]{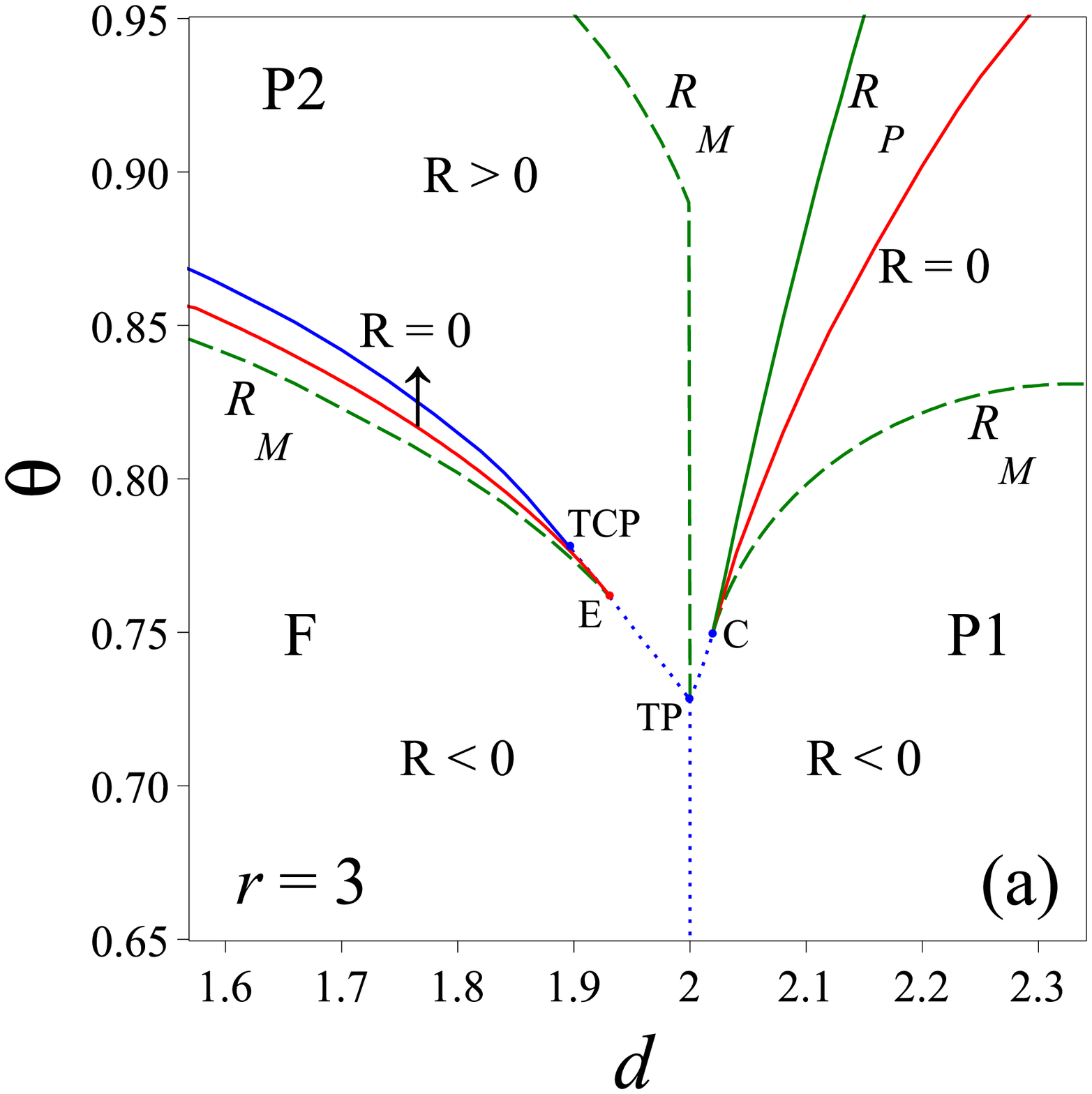}
	\includegraphics[width=0.45\linewidth]{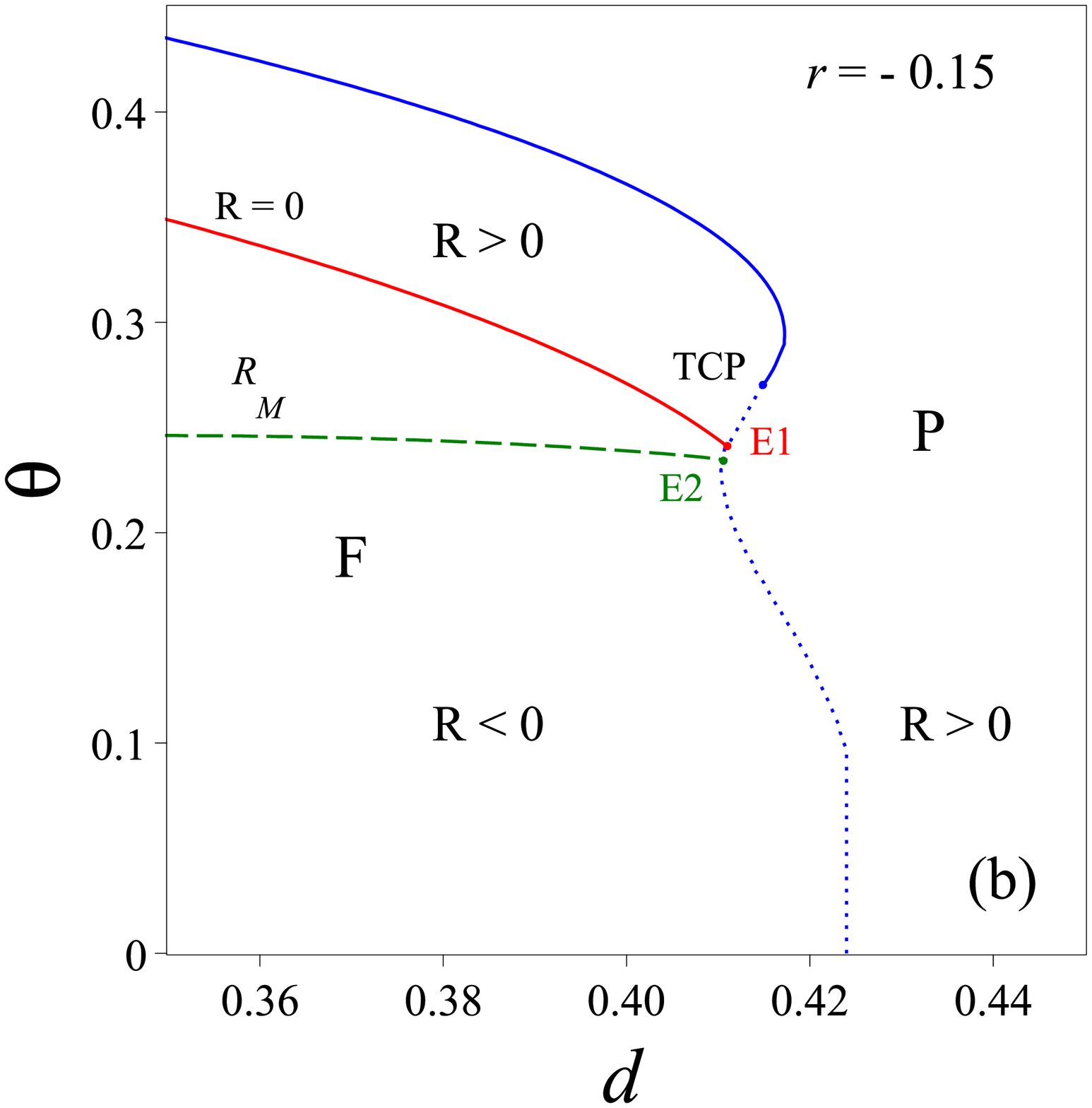}
	\includegraphics[width=0.45\linewidth]{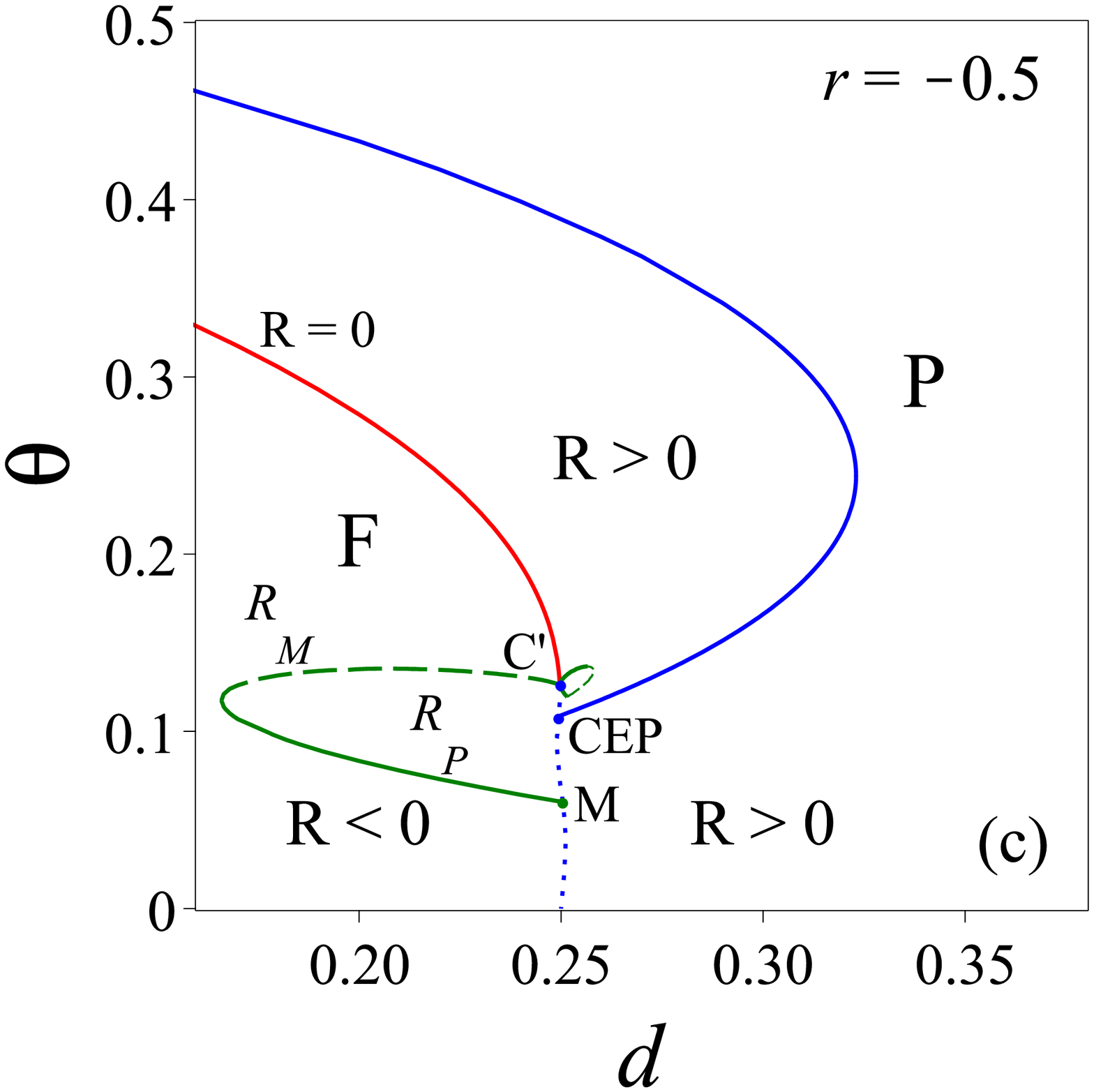}
	\includegraphics[width=0.45\linewidth]{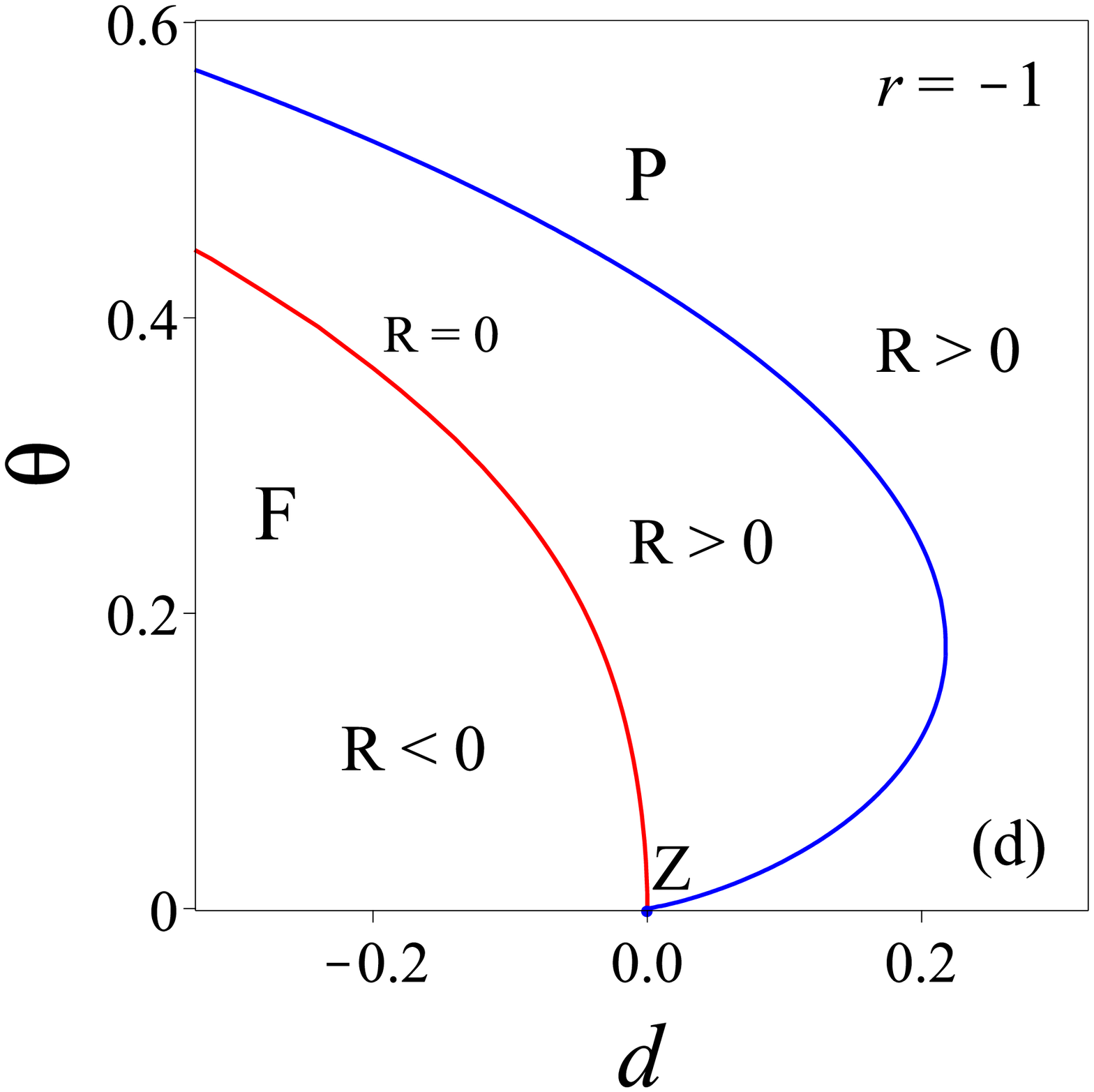}
	\caption{Geometrical phase diagrams with a $R=0$ boundary line for several values of $r$ in the $d-\theta$ plane when $h=0$. {\textbf{a}} $r=3$, {\textbf{b}} $r=-0.15$, {\textbf{c}} $r=-0.5$, {\textbf{d}} $r=-1$. Blue dotted and blue solid lines represent the first-order and second-order phase transitions, respectively. Red solid lines corresponds to $R=0$, green dashed lines corresponds to at which $R$ minimum and green solid lines corresponds to at which $R$ maximum.} \label{fig2}
\end{figure*} Clearly, a geometric phase transition occurs in the F and P1 phases. In Fig. 3b, as in Fig. 3a, $R=0$ and $R_M$ curves lie down under the second-order phase transition line. But, different from Fig. 3a, they terminate at two distinct points on the first-order phase transition curve denoted by E1 ($\theta_{E1}=0.241,d_{E1}=0.411$) and E2 ($\theta_{E2}=0.234,d_{E2}=0.411$). On the other hand, in Fig. 3c, while an $R_P$ curve occurs that forms a half ring and completes the $R_M$ curve, $R=0$ and $R_M$ curves meet at the internal critical point C$'$, in the F phase. $R_P$ curve ends up at the M point ($\theta_M=0.060, d_M=0.250$) which is under the CEP over the first-order phase transition line. Again, in the F phase, in the region that is between $R=0$ line and second-order phase transition line, $R_P$ and $R_M$ curves constitute a small ring which starts on C$'$ and ends up on the same point. Finally, different from the other GPDs the geometric phase diagram in Fig. 1d has only one curve (or $R=0$ curve) in the F phase and that curve ends up at the zero temperature critical point Z. Hence, $R=0$ curves separate $R<0$ and $R>0$ regions in the F phase in the GPDs. It should be noted that, $R>0$ regions in F phase get large with decreasing $r$ if $r<0$ although it is too small when $r>0$. On the other hand, always $R>0$ in all paramagnetic phases except P1.

\section{Conclusions}

In conclusion, we presented in this paper within a thermodynamic geometry approach called Ruppeiner geometry some graphical results for the Ricci scalar $R$ from which a deeper insight into the microscopic interactions of spins can be observed in the anisotropic BEG model. We have determined the locations of $R=0, R_P$ and $R_M$ lines in some of multicritical phase diagrams drawn previously in \cite{[4], [5]}. One of the novel results observed in the diagrams is that there exist for $r=3$ two $R = 0$ curves one of which prolongates beyond the external critical point C and separates the $R<0$ region in P1 and $R>0$ regime in P2 while the other curve is located in the F phase  as in \cite{[17]}. As for the extrema in $R$, single $R_{P}$ line appears as another prolongation line beginning from C towards P2 phase. But, we observed multiple number of $R_{M}$ lines each of which is located in a different phase of the same multicritical topology. For example, the minimum curves lying in the P1 and P2 phases start at C and TP, respectively. On the other hand, there are differences in the number and locations of above curvature lines as far as the reentrence phenomenon with $r<0$ is concerned, since $R=0$ curve divides the F phase into two large zones with opposite curvature signs and joins with $R_{M}$ line at internal critical point C$'$ if $r=-0.5$ or directly reaches to zero-temperature point Z when $r=-1$ in F phase. Hence, important roles of TP, C, C$'$, CEP and Z in the GPDs become appearent which can also be observed through the analysis of singularities (with a sign change) for the Ricci scalar. Another novel finding is the terminations of $R=0$ and $R_{M}$ lines at two separate points on the F-P coexistence curve for $r=-0.15$, not seen in \cite{[17]}.

{\bf Data Availability Statement} No Data associated in the manuscript.

\end{document}